\documentclass[twocolumn,twoside,preprintnumbers,amsmath,amssymb,showkeys]
{revtex4}
\usepackage{epsfig}
\usepackage{graphicx}

\usepackage{fancyhdr}

\usepackage{pslatex}

\def\beq{\begin{equation}}
\def\enq{\end{equation}}
\def\beqa{\begin{eqnarray}}
\def\enqa{\end{eqnarray}}
\def\MeV{\nobreak\,\mbox{MeV}}
\def\GeV{\nobreak\,\mbox{GeV}}

\def\pli{p^\prime}
\def\la{\lambda}
\def\qq{\lag\bar{q}q\rag}
\def\ss{\lag\bar{s}s\rag}
\newcommand{\rag}{\rangle}
\newcommand{\lag}{\langle}
\def\lb{\label}
\def\nn{\nonumber}

\begin{document}

\title{QCD sum rule approach to the new mesons and the $g_{D_{sJ}DK}$
coupling constant}

\author{Marina Nielsen}

\affiliation{Instituto de F\'{\i}sica, Universidade de S\~ao Paulo, 
         C.P. 66318, 05315-970 S\~ao Paulo-SP, Brazil}
         

\begin{abstract}
We use  diquark-antidiquark currents to investigate the masses and partial
decay widths  of the recently observed mesons 
$D_{sJ}^{+}(2317)$, $D_0^{*0}(2308)$ and $X(3872)$, considered as four-quark 
states, in a QCD sum rule approach. In particular we investigate the
coupling constant $g_{D_{sJ}DK}$. We found that the $g_{D_{sJ}DK}$ obtained
in this four-quark scenario is smaller than the coupling constant obtained
when $D_{sJ}^{+}(2317)$ is considered as a conventional $c\bar{s}$ state.


\end{abstract}
\maketitle

\thispagestyle{fancy}

\setcounter{page}{1}

\section{Introduction}

The constituent quark model provides a rather successful descrition of the
spectrum of the mesons in terms of quark-antiquark bound states, which fit
into the suitable multiplets reasonably well. Therefore, it is understandable
that the recent observations of the very narrow resonances
$D_{sJ}^+(2317)$ by BaBar \cite{babar},
$D_{sJ}^{+}(2460)$ by CLEO \cite{cleo}, $X(3872)$ by BELLE \cite{BELLE},
and the very broad scalar meson $D_0^{*0}(2308)$ by BELLE \cite{belle2},
all of them with masses below quark model predictions, have stimulated a
renewed interest in the spectroscopy of open charm and charmonium states.
The difficulties to identify the mesons $D_{sJ}^+(2317)$ and $D_{sJ}^{+}(2460)$
as $c\bar{s}$ states are rather similar to those appearing in the light
scalar mesons below 1 GeV (the isoscalars $\sigma(500),~f_0(980)$, the 
isodublet $\kappa(800)$ and the isovector $a_0(980)$), that can be interpreted
as four-quark states  \cite{jaffe,cloto}. In the case of $X(3872)$, besides 
its small mass, the observation, reported by the BELLE collaboration 
\cite{belleE}, 
that the $X$ decays to $J/\psi\,\pi^+\pi^-\pi^0$, with a strength that
is compatible to that of the  $J/\psi\pi^+\pi^-$ mode:
\beq
{Br(X\to J/\psi\,\pi^+\pi^-\pi^0)\over Br(X\to J/\psi\,\pi^+\pi^-)}
=1.0\pm0.4\pm0.3\;,
\label{data}
\enq 
establishes strong isospin violating effects, which can not be explained
if the $X(3872)$ is interpreted as a $c\bar{c}$ state.
 
Due to these facts,
these new mesons were considered as good candidates for  four-quark states  
by many authors \cite{Swanson}. 
In refs.~\cite{blmnn,x3872} the  method of QCD  sum rules (QCDSR) 
\cite{svz,rry,narison} was used to study the two-point functions for the mesons
$D_{sJ}^+(2317)$, $D_0^{*0}(2308)$and $X(3872)$ considered as four-quark 
states in a  diquark-antidiquark configuration.
The results obtained for their masses are compatible with the experimental
values and, therefore, in refs.~\cite{blmnn,x3872} the authors concluded
that it is possible to reproduce 
the experimental value of the masses using a four-quark representation for 
these states. 

Concerning their decay widths, the study of the three-point functions 
related to the decay widths  $D_{sJ}^+(2317)\to D_s^+\pi^0$,
$D_0^{*0}\to D^+\pi^-$  and 
$X(3872)\to J\psi\pi^+\pi^-$,
using the  diquark-antidiquark configuration for $D_{sJ}$, $D_0^{*0}$ and 
$X$, was done in refs.~\cite{decayds,decayd0,decayx}.  The results obtained 
for their partial decay widths are given in Table I, from where 
we see that the partial decay widths obtained in 
refs.~\cite{decayds,decayd0}, supposing that the mesons  $D_{sJ}^+(2317)$ 
and $D_0^{*0}$ are 
four-quark states, are consistent with the experimental upper information for 
the total decay width.

\begin{table}[htb]
\begin{center}
\caption{Numerical results for the resonance decay widths}
\end{center}
\begin{tabular}{|c|c|c|c|} 
\hline
decay & $D_{sJ}^+\to D_s^+\pi^0$ & $D_0^{*0}\to D^+\pi^-$ & $X\to J/\psi
\pi^+\pi^-$   \\
\hline
$\Gamma$ (MeV) &$(6\pm2)\times10^{-3}$  & $120\pm20$ & $50\pm15$\\
\hline
$\Gamma^{exp}_{tot}$ (MeV) &$<5$  & $270$ & $<2.3$\\
\hline
\end{tabular}
\label{tab1}
\end{table}
However,
in the case of the meson $X(3872)$, the  partial decay width obtained in 
ref.~\cite{decayx} is much bigger than the experimental upper limit to the
 total width. 

In ref.~\cite{decayx} some arguments were presented to reduce the value
of the $X(3872)$ decay width, by imposing that the initial four-quark state
needs to have a non-trivial color structure. In this case, its partial decay 
width can be reduced to
$\Gamma(X\to J/\psi\pi^+\pi^-))=(0.7\pm0.2)~\MeV$. However,
that procedure may appear somewhat unjustified and, therefore, more study
is needed until one can arrive at a definitive conclusion about the 
structure of the meson $X(3872)$.

Concerning the meson $D_{sJ}^+(2317)$, although its mass and decay width can be
explained in a four-quark scenario, they can also be reproduced in 
other approaches \cite{Swanson}, and it is not yet
possible to discriminate between the different structures proposed for this 
state. Therefore, it is important to find experimental observations
that could be used to descriminate between the different quark structure of 
these mesons. As pointed out in ref.~\cite{polosa}, a signal could be obtained
by the analysis of certain heavy-ion collision observables. In particular,
the meson $D_{sJ}^+(2317)$ can be produced in reactions induced by photons
on kaon targets in a nuclear medium formed in a heavy-ion collision.
Therefore, if the coupling constant, $g_{D_{sJ}DK}$, is found to be very 
different depending on the structure for $D_{sJ}^+(2317)$, then the
photo-production of $D_{sJ}^+(2317)$ can be used as a signal to descriminate
its structure.

\section{The $g_{D_{sJ}DK}$ coupling constant}

The coupling, $g_{D_{sJ}DK}$, supposing that the meson $D_{sJ}^+(2317)$
is a conventional $c\bar{s}$ state, was evaluated in ref.~\cite{ww}. They
got:
\beq
g_{D_{sJ}DK}=(9.2\pm0.5)~\GeV
\label{gww}
\enq 
Here, we extend the calculation done in refs.~\cite{decayds,decayd0} to study 
the hadronic vertex $D_{sJ} DK$.
The QCDSR calculation for the vertex, $D_{sJ}DK$, centers
around the three-point function given by

\beq
T_\mu(p,\pli,q)=\int d^4x d^4y ~e^{i\pli.x}~e^{iq.y}
\lag 0 |T[j_{D}(x)j_{5\mu}(y)j^\dagger_{D_{sJ}}(0)]|0\rag,
\lb{3po}
\enq
where $j_{D_{sJ}}$ is the interpolating field for the scalar $D_{sJ}$ meson
\cite{blmnn}:
\beq
j_{D_{sJ}}={\epsilon_{abc}\epsilon_{dec}\over\sqrt{2}}
\left[(u_a^TC\gamma_5c_b)
(\bar{u}_d\gamma_5C\bar{s}_e^T)+u\leftrightarrow d\right],
\label{int}
\enq
where $a,~b,~c,~...$ are colour indices and $C$ is the charge conjugation
matrix. 
In Eq.~(\ref{3po}), $p=\pli+q$ and the interpolating fields for the kaon
and for the $D$ mesons are given by:
\beq
j_{5\mu}=\bar{s}_a\gamma_\mu\gamma_5q_a
,\,\;\;\;j_{D}=i\bar{q}_a\gamma_5c_a,
\lb{pseu}
\enq
where $q$ stands for the light quark $u$ or $d$.

The calculation of the phenomenological side proceeds by 
inserting intermediate states for $D$, $K$ and $D_{sJ}$, and by using 
the definitions: 
$\lag 0 | j_{5\mu}|K(q)\rag =iq_\mu F_{K}$,
$\lag 0 | j_{D}|D(\pli)\rag ={m_{D}^2f_{D}\over m_c}$,
$\lag 0 | j_{D_{sJ}}(p)\rag =\la$.
We obtain the following relation:
\beqa
T_{\mu}^{phen} (p,\pli,q)&=&{\la m_{D}^2f_{D}F_{K}~
g_{D_{sJ}DK}/m_c
\over (p^2-m_{D_{sJ}}^2)({\pli}^2-m_{D}^2)(q^2-m_K^2)}~
~q_\mu \nn\\
&+&\mbox{continuum contribution}\;,
\lb{phen}
\enqa
where the coupling constant, $g_{D_{sJ}DK}$, is defined by the 
on-mass-shell matrix element: $\lag D K|D_{sJ}\rag=g_{D_{sJ}DK}$.
The continuum contribution in Eq.(\ref{phen}) contains the contributions of
all possible excited states.

In the case of the light scalar mesons, considered as diquark-antidiquark 
states, the study of their vertices functions using the QCD sum rule approach 
at the pion pole \cite{narison,rry,nari2}, was done in ref.\cite{sca}. It was
shown that  the decay widths determined from the QCD sum rule calculation are 
consistent with existing experimental data. 
Here, we follow ref.~\cite{bracco} and work at the kaon pole. 
The main reason for working at the kaon pole is that  one does not 
have to deal with the complications associated with the extrapolation of the 
form factor \cite{dosch}. The kaon pole method consists in neglecting the kaon 
mass in the denominator of Eq.~(\ref{phen}) and working at $q^2=0$. In the 
OPE side one singles out the leading terms in the operator product expansion 
of Eq.(\ref{3po}) that match the $1/q^2$ term. Since we are working at 
$q^2=0$, we take the limit $p^2={\pli}^2$ and we 
apply a single Borel transformation to $p^2,{\pli}^2\rightarrow M^2$.
In the phenomenological side, in the structure $q_\mu$ we get:
\beqa
T^{phen}(M^2)&=&
{\la m_{D}^2f_{D}F_{K}~
g_{D_{sJ}DK}\over m_c(m_{D_{sJ}}^2-m_{D}^2)}\left(
e^{-m_{D}^2/M^2} -e^{-m_{D_{sJ}}^2/M^2}\right)\nn\\
&+&A~e^{-s_0/M^2}+
\int_{u_0}^\infty\rho_{cc}(u)~e^{-u/M^2}du,
\label{paco}
\enqa
where $A$ and $\rho_{cc}(u)$
stands for the pole-continuum transitions and pure continuum contributions,
with $s_0$ and $u_0$ being the continuum thresholds for $D_{sJ}$ and $D$ 
respectively \cite{decayds,decayd0}. 
For simplicity, one assumes that the pure continuum 
contribution to the spectral density, $\rho_{cc}(u)$, is given by the result 
obtained in the OPE side. 
Therefore, one uses the ansatz: $\rho_{cc}(u)=\rho_{OPE}(u)$.
In Eq.(\ref{paco}), $A$ is a parameter which, together with
$g_{D_{sJ}DK}$, has to be determined by the sum rule.

In the OPE side we single out the leading terms proportional to
$q_\mu/q^2$.  Transferring the pure continuum contribution to
the OPE side, the sum rule for the coupling constant, up to dimension 7, is 
given by:
\beqa
&&C~\left(e^{-m_{D}^2/M^2} -e^{-m_{D_{sJ}}^2/M^2}\right)+A~e^{-s_0/M^2}=
\nn\\
&=&{1\over\sqrt{2}}\left[{\qq+\ss\over2^4\pi^2}\int_{m_c^2}^{u_0}du~e^{-u/M^2}
u\left(1-{m_c^2\over u}\right)^2\right.\nn\\
&+&{m_cm_s\ss\over2^5\pi^2}\int_{m_c^2}^{u_0}du~e^{-u/M^2}
\left(1-{m_c^2\over u}\right)^2+{m_s\qq\ss\over12}e^{-m_c^2/M^2}
\nn\\
&-&\left.{m_c\qq(\qq+\ss)\over3}e^{-m_c^2/M^2}\right],
\label{sr}
\enqa
with 
\beq
C={\la m_{D}^2f_{D}F_{K}
\over m_c(m_{D_{sJ}}^2-m_{D}^2)}~g_{D_{sJ}DK}.
\label{coef}
\enq

\section{Results and conclusions}

In the numerical analysis of the sum rules, the values used for the meson
masses, quark masses and condensates are: $m_{D_{sJ}}=2.317~\GeV$,
$m_{D}=1.87~\GeV$, $m_c=1.2\,\GeV$, $m_s=.13\,\GeV$ 
$\lag\bar{q}q\rag=\,-(0.23)^3\,\GeV^3$, $\lag\bar{s}s\rag=0.8\qq$.
For the meson
decay constants we use $F_K=160~\MeV$ and $f_{D}=0.22~\GeV$
\cite{cleo2}. We use $u_0=6~\GeV^2$ and
for the current meson coupling, $\la$, we are going
 to use the result obtained from the two-point function in ref.~\cite{blmnn}.
Considering $2.6\leq\sqrt{s_0}\leq2.8~\GeV$ we get $\la=(2.9\pm0.3)\times
10^{-3}~\GeV^5$.

\begin{figure}[!htb]
\begin{center}
\includegraphics*[width=7.6cm]{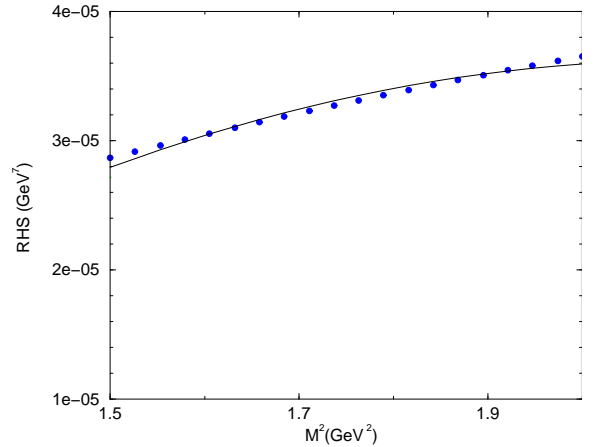} 
\end{center}
\vspace*{-.8cm}
\caption{Dots: the RHS of Eq.(\ref{sr}), as a function of the Borel 
mass.  The solid line gives the fit of the QCDSR results through 
the LHS of Eq.(\ref{sr}).}
\label{figure}
\end{figure} 

In Fig.~1 we show, through the dots, the right-hand side (RHS) of 
Eq.(\ref{sr}) as a function of the Borel mass.
To determine $g_{D_{sJ}DK}$ we fit the QCDSR results with the analytical
expression in the left-hand side (LHS) of Eq.(\ref{sr}):
\beq
C\left(
e^{-m_{D_s}^2/M^2} -e^{-m_{D_{sJ}}^2/M^2}\right)+A~e^{-s_0/M^2},
\label{exp}
\enq
Using $\sqrt{s_0}=2.7\GeV$ we get: $C=4.53\times10^{-4}~\GeV^7$ and 
$A=-4.68\times10^{-4}~\GeV^7$. Using the definition of $C$ in Eq.(\ref{coef})
and $\la=2.9\times10^{-3}~\GeV^5$ (the value obtained for $\sqrt{s_0}=2.7
\GeV$) we get $g_{D_{sJ}DK}=2.8~\GeV$. Allowing $s_0$ to vary in the
interval $2.6\leq\sqrt{s_0}\leq2.8~\GeV$, the corresponding variation
obtained for the coupling constant is 
\beq
2.5~\GeV\leq g_{D_{sJ}DK}\leq 3.8~\GeV.
\label{g}
\enq

Fixing $\sqrt{s_0}=2.7\GeV$ and varying the quark condensate, the charm quark 
and the strange quark masses in the intervals: $-(0.24)^3\leq\lag\bar{q}q\rag
\leq-(0.22)^3\,\GeV^3$, $1.1\leq m_c\leq 1.3\GeV$ and 
$0.11\leq m_s\leq 0.15\GeV$, we get results for the coupling constant
still between the lower and upper limits given above. it is important
to mention that the agreement between the RHS and LHS of the sum rule
in Fig.1 is not so good, in this case, as it was in the case of the
couplings $g_{D_{sJ}D_s\pi}$ and $g_{D_0^{*0}D\pi}$ evaluated in 
refs.~\cite{decayds,decayd0}. One possible reason for that is the fact that
the kaon mass is much bigger than the pion mass. Therefore, neglecting 
the kaon mass in Eq.~(\ref{phen}) is not an approximation as good as it is 
in the case of the sum rule in the pion pole.

We have presented a QCD sum rule study of the vertex function associated with
the hadronic vertex $D_{sJ}DK$, where the
$D_{sJ}(2317)$ meson was considered as diquark-antidiquark state. 
Comparing the results in Eqs.~(\ref{g}) and (\ref{gww}) we see that when the
meson $D_{sJ}(2317)$ is considered as a conventional $c\bar{s}$ state one
gets a $g_{D_{sJ}DK}$ coupling constant much bigger than when $D_{sJ}(2317)$
is considered a four-quark state. This result can be usefull to experimentally
investigate the quark structure of the meson  $D_{sJ}(2317)$ through its
photon production in a nuclear medium.

\noindent{\bf Acknowledgements} 

This work has been supported by CNPq and FAPESP.


\end{document}